\documentclass[final,5p,times,twocolumn]{elsarticle}

 \usepackage{graphics}

\usepackage{amssymb}

\begin{document}

\begin{frontmatter}



\title{Transformation between statistical ensembles in the modelling of nuclear fragmentation}


\author{G. Chaudhuri$^1$, F. Gulminelli$^2$, S.Mallik$^1$}

\address{$^1$Theoretical Physics Division, Variable Energy Cyclotron Centre, 1/AF Bidhan Nagar, Kolkata700064,India}

\address{$^2$LPC Caen IN2P3-CNRS/EnsiCaen et Universite, Caen, France}
\begin{abstract}
We explore the conditions under which the particle number conservation constraint deforms the
predictions of fragmentation observables as calculated in the grand-canonical ensemble.
We derive an analytical formula allowing to extract canonical results from a grand-canonical calculation and vice-versa. This formula  shows that exact canonical results can be recovered for observables varying linearly or quadratically
 with the number of particles, independent of the grand-canonical particle number fluctuations.
We explore the validity of such grandcanonical extrapolation for different fragmentation
observables in the framework of the analytical Grand Canonical or Canonical Thermodynamical Model [(G)CTM] of nuclear multifragmentation.
It is found that corrections to the grandcanonical expectations can be evaluated with high precision,
provided the system does not experience a first-order phase transition.
 In particular, because of the Coulomb quenching of the
liquid-gas phase transition of nuclear matter, we find that mass conservation
 corrections to the grandcanonical ensemble can be safely computed  for typical observables of interest in experimental measurements of nuclear fragmentation, even if deviations exist for highly exclusive observables.
\end{abstract}


\end{frontmatter}
\section{Introduction}
Statistical ensembles are known to give different predictions in finite systems, and  to converge at the thermodynamical limit if interactions are short-range. The inequivalence is particularly pronounced, and the convergence correspondingly slow, in the presence of phase transitions\cite{gross,Herzog}.
In the context of nuclear physics,  the conditions of how this convergence is realized by increasing the particle number  can be studied  taking the model case of a neutral system\cite{gargi07}. Neutral nuclear
systems  do not exist in nature, but this simplification  is often introduced in the context of nuclear matter.
In the framework of nuclear models with cluster degrees of freedom,  a finite counterpart to nuclear matter can be realized in practice by switching off  the Coulomb interactions both in the cluster energy functional and in the inter-cluster interactions, and allowing any arbitrary cluster size in the statistical equilibrium. Such a system exhibits a first order liquid-gas phase transition at the thermodynamic limit, which makes ensembles strongly inequivalent for any finite size $N$\cite{gargi07,fisher}.

Another interesting case is obtained considering the statistical equilibrium of charged nuclear fragments, as it might be realized in heavy-ion collisions. In this case the thermodynamic limit is not defined, since the energy density diverges for $N, V\to\infty$. However a thermodynamic limit can be defined again if we take the physical case of neutron star crusts, where clusters are charged, but the overall charge is screened by a uniform electron background. This case is studied in detail in ref.\cite{inequivalence}, and again strong signatures of ensemble inequivalence are seen.
An intermediate case is found in a model case where no uniform electrons are considered, but the cluster size is artificially set to a finite value $A_{max}$. This model is analyzed in detail in ref.\cite{gargi07}. A thermodynamic limit can be set for such a model, and the convergence between ensembles is the faster the smaller is $A_{max}$.
This can be understood from the fact that the liquid-gas phase transition present in the uncharged model is quenched in that case, since no liquid fraction (which corresponds to  $A_{max}\to\infty$) is allowed in the model. The physical meaning of such a model at the thermodynamic limit is not very clear. However if we concentrate on finite systems only, this model has some relevance in the study of nuclear multifragmentation, where a maximum size is imposed by the repulsive coulomb interaction, and there is no electron background to neutralize it.  The results of ref.\cite{gargi07} thus suggest that, if we consider finite systems only, ensembles may be close to equivalence. This is confirmed in ref.\cite{gargi12}, where it is shown that typical inclusive fragmentation variables converge if temperature is not too low and multiplicity is sufficiently high to avoid important finite number effects.

Motivated by these works, in this paper we concentrate on finite charged nuclear systems without any electron screening,   described in the framework of a statistical model with cluster degrees of freedom, and work out analytical relations connecting the different statistical ensembles.
We propose an approximate expression allowing to transform the observables from one ensemble to the other.
In particular, we  show that the modification of grand canonical results due to particle number conservation can be exactly computed  for observables varying linearly or quadratically with the number of particles, even in the regime of very small systems where particle number fluctuations in the grand canonical ensemble cannot be neglected.

In the more general case, transformations are not exact and the quality of the transformation formula depends not only on the system size and temperature, but also on the specific model. Most models in statistical mechanics cannot be solved analytically in any arbitrary ensemble, meaning that it is difficult to control the quality of the approximation. However,
  the well-known  Grand Canonical or Canonical Thermodynamical Model [(G)CTM] of nuclear multifragmentation has the advantage of being analytically solvable both in the canonical and in the grandcanonical ensemble.
Comparing the analytical extrapolation of the grand canonical ensemble to the exact resolution of the canonical one, we will then be able to see how well one can account for a conservation law (here: particle number) without explicitely calculating the corresponding partition sum (here: the canonical one).

\section{Transformation between statistical ensembles}
There are two ways of computing the grandcanonical average number of particles $<N>$ at fugacity $\alpha=\beta\mu$,
where $\beta$ is the inverse temperature and $\mu$ the chemical potential. The first way needs the calculation of the grand-canonical partition sum $Z_\alpha$
\begin{equation}
<N>_\alpha=\frac{\partial ln Z_\alpha}{\partial \alpha}
\end{equation}
While the second way uses the definition of the particle number distribution in the grandcanonical ensemble
\begin{equation}
<N>_\alpha=\sum_{N=0}^{\infty} N P_\alpha (N)
\end{equation}
This distribution is given by
\begin{equation}
 P_\alpha (N)=Z_{\alpha}^{-1} Z_N \exp \alpha N \label{proba}
\end{equation}
and implies the knowledge of the canonical partition sum.
Note that the knowledge of $Z_{\alpha}$ is not really necessary in this last equation because it can be deduced by the condition of normalization of probabilities.
The same kind of relations is established for the particle variance:
\begin{equation}
\sigma^2_\alpha=\frac{\partial^2 ln Z_\alpha}{\partial \alpha^2}=
\sum_{N=0}^{\infty} \left (N-<N>_\alpha \right )^2  P_\alpha (N) \label{eq2}
\end{equation}

This analytical connection between canonical and grandcanonical suggests that we should be able to extract grandcanonical results from canonical ones and viceversa, provided the probability distribution
is completely described by a limited number of moments. This is particularly true if this distribution is a gaussian (defined only by mean value and variance) as we now show.

 Let us consider a given inverse temperature $\beta$ and a given volume $V$ which we will suppose fixed and omit from all the notations.  Quantities calculated in the grandcanonical ensemble will have the suffix "GC", quantities calculated in the canonical ensemble will be noted "C".
We will concentrate on  a generic observable of interest $Q$  which can be computed either in the canonical ($Q_C$) or in the grandcanonical ($Q_{GC}$) ensemble.
Starting from the exact relation connecting canonical and grandcanonical:

\begin{equation}
Q_{GC}(\alpha)=\sum_N P_\alpha (N) Q_C(N)
\label{eq3}
\end{equation}

we do a Taylor developement of $Q_C(N)$ around $N=N_{GC}(\alpha)$ truncated at second order:

\begin{eqnarray}
Q_{C}(N) &\approx& Q_{C}(N=N_{GC}) \nonumber \\
&+&  (N-N_{GC})  \frac{\partial Q_{C}}{\partial N}(N=N_{GC})  \nonumber \\
&+&
\frac 12  (N-N_{GC})^2  \frac{\partial^2 Q_{C}}{\partial N^2}(N=N_{GC})
\label{eq4}
\end{eqnarray}

We replace in eq.(\ref{eq3}):

\begin{equation}
Q_{GC}(\alpha) \approx Q_{C}(N_{GC}) +\frac 12  \sigma^2_{GC}(\alpha) \frac{\partial^2 Q_{C}}{\partial N^2}|_{N=N_{GC}}
\label{eq5}
\end{equation}

where $\sigma^2_{GC}(\alpha)$ is given by eq.(\ref{eq2}).

This result indicates that the difference between  the two predictions does not only increase with increasing particle number fluctuation (which is linked to the system size and temperature, and independent of the observable $Q$),{ but also with increasing convexity of the observable}\cite{maras}.

From a technical point of view, it is always much simpler to calculate an observable in the grandcanonical ensemble than in the canonical one. On the other side, in realistic modelling of nuclear fragmentation, the correct ensemble is rather the canonical or the microcanonical one. Indeed nuclear systems that can be formed in the laboratory are isolated systems which are not coupled to an external energy and particle bath. The excited nuclear  sources which can be described via statistical models typically constitute only a subsystem of the total interacting system, meaning that conservation laws on particle number and energy are not strict. However, energy and particle numbers can be in principle measured, and statistical ensembles with a fixed number of particles and energy can be obtained by an appropriate sorting of experimental data.
Therefore an expression similar to eq.(\ref{eq5}), but which would express the microcanonical, or at least the canonical result as a function of the grandcanonical one, would be most welcome. We leave the extension of the formalism to the implementation of energy conservation to a future work, and concentrate here on the mass conservation constraint.
Let us call $C(N)= \frac{\partial^2 Q_{C}}{\partial N^2}(N)$.
The Taylor expansion eq.(\ref{eq4}) gives:

\begin{equation}
C(N) \approx C(N_{GC}) +\  (N-N_{GC})  \frac{\partial C}{\partial N}+
\frac 12  (N-N_{GC})^2  \frac{\partial^2 C}{\partial N^2}
\label{eq6}
\end{equation}

and the grandcanonical estimation of $C$ is given by

\begin{equation}
C_{GC} \approx C(N_{GC}) +\frac 12  \sigma^2_{GC}(\alpha) \frac{\partial^2 C}{\partial N^2}|_{N=N_{GC}}
\label{eq7}
\end{equation}

We replace eq.(\ref{eq7}) into eq.(\ref{eq5}) and consider the limit of small particle number fluctuations,
$ \sigma^2_{GC}/N_{GC}^2 < 1$. This limit is not realized in phase transitions, but otherwise it should be correct.
Within this limit we can neglect terms of the order of $\sigma^4_{GC}$ and we get:

\begin{equation}
Q_{C}(N_{GC}) \approx Q_{GC}(\alpha)   -\frac 12  \sigma^2_{GC}(\alpha) \frac{\partial^2 Q_{GC}}{\partial N_{GC}^2}|_{\alpha=\alpha(N_{GC})}
\label{eq8}
\end{equation}

This is the desired expression, since the r.h.s. of this formula can be entirely calculated in the grandcanonical ensemble.

\section{The fragmentation model}

Let us take the (Grand) Canonical Thermodynamical Model [(G)CTM] as defined in ref.\cite{gargi07}. If isospin degrees of freedom are taken into account,
the canonical partition sum depends on the two independent variables given by the total neutron and proton number. To simplify the discussion we consider in this work a single particle number variable $N$. We attribute to each cluster of size A an effective charge $Z_{eff}=A/2$ and introduce the Coulomb interaction in the Wigner-Seitz approximation\cite{dasgupta}. This corresponds to an approximate treatment of isospin symmetric matter, and the extension to isospin asymmetry is left for a future work.

The canonical partition sum  is given by
\begin{equation}
Z_N=\sum_{\vec{n}:N} \prod_{A=1}^{A_{max}} \frac{\omega_A^{n_A}}{n_A!}\label{zcano}
\end{equation}
where $A_{max}$ is the maximum allowed size and the sum comprises all channels $\vec{n}=\{n_1,\dots,n_{A_{max}}\}$
such that $N=\sum_{A=1}^{A_{max}} n_A A$. $\omega_A$ is the partition sum of a cluster of size $A$, given by\cite{dasgupta}:
\begin{eqnarray}
\omega_A&=&\frac{V_{free}}{h^3} \left ( \frac{2\pi m A}{\beta} \right )^{3/2} \nonumber \\
&\cdot& \exp\left [ -\beta \left ( W_0 A -\sigma(\beta) A^{2/3} +A/(\epsilon_0 \beta^2) \right ) \right ] ,
\end{eqnarray}
where  $V_0(N)$ is the normal volume of a nucleus composed of $N$ nucleons, $V_{free}=V-V_0(N)$ is the free volume,
 $m$ the nucleon mass, and $W_0, \sigma, \epsilon_0$ are parameters\cite{dasgupta}.
The partition sum eq.(\ref{zcano}) can be calculated using a recursion relation\cite{dasgupta}:
\begin{equation}
Z_N= \frac{1}{N}\sum_{A=1}^N A \omega_A  Z_{N-A} .
\label{zcan}
\end{equation}
This expression can be recursively computed with the initial condition
$Z_1=\omega_1$.
Let us take the standard statistical definition of the grandcanonical ensemble as
\begin{equation}
Z_{\alpha}=\sum_{N=0}^\infty Z_N \exp \alpha N \label{stat}
\end{equation}
Replacing the SMM expression for the canonical partition sum we g,et
\begin{equation}
Z_{\alpha}=\sum_{N=0}^\infty \sum_{\vec{n}:N} \prod_{A=1}^{A_{max}} \frac{\omega_A^{n_A}}{n_A!}\exp \alpha N
\end{equation}
which can be rewritten as
\begin{eqnarray}
Z_{\alpha}&=&\sum_{n_1=0}^\infty \frac{\omega_1^{n_1}}{n_1!}  \dots \sum_{n_ {A_{max}}=0}^\infty \frac{\omega_{A_{max}}^{n_{A_{max}}}}{n_{A_{max}}!} \exp \alpha N  \nonumber\\
&=&\prod_{i=1}^{A_{max}}\sum_{n_i=0}^\infty  \frac{\omega_i^{n_i}}{n_i!}\exp \left ( \alpha \sum_{A=1}^{A_{max}} n_A A \right ) \nonumber \\
&=& \prod_{A=1}^{A_{max}}\sum_{n_A=0}^\infty  \frac{ \left( \omega_A
\exp\alpha_A \right )^{n_A}}{n_A!} \nonumber\\
&=&  \prod_{A=1}^{A_{max}} \exp \left ( \omega_A
\exp\beta\mu_A \right )
\end{eqnarray}
which is nothing but the standard expression of the SMM grandcanonical partition sum\cite{dasgupta}  $Z_{GC}$ (where $\alpha_A=\alpha\cdot A$).
This shows that the canonical and grandcanonical partition sums  satisfy the general relation eq.(\ref{stat}), and the quantity $P_\alpha (N)$ defined by equation (\ref{proba}) can indeed be interpreted as a probability.

It is important to remark that in order to have the correct mapping between canonical and grandcanonical
eq.(\ref{stat}) the vacuum canonical partition sum $Z_0$ has to be considered. This quantity
is not defined by the recursion relation, however we can extract it  from the probability normalization condition:
%
%

\begin{equation}
 P_\alpha (0)=1-  \frac{1}{Z_{\alpha}} \sum_{N=1}^\infty Z_N \exp \alpha N
\end{equation}
leading to:
\begin{equation}
 Z_0=P_\alpha (0)Z_{\alpha}
\end{equation}
This discussion might sound academic but this is not entirely so. Indeed in the case of first order phase transitions we might have a bimodality in the grandcanonical particle number distribution, which reflects the equilibrium which would be obtained at the thermodynamic limit between a dilute (small number of particles) and a dense (high number of particles) phase.
It is well known from mean-field studies that this phase equilibrium might be obtained mixing a finite density phase with the vacuum. This is notably the case for the nuclear liquid-gas phase transition at zero temperature (see for instance ref.\cite{ducoin}). In these situations, it is important to have the correct grandcanonical weight for the vacuum solution, even if of course from the canonical point of view the thermodynamics of the vacuum has no interest.

\section{Results}

Since the thermodynamic fragmentation model is exactly solvable both in the canonical and in the grand canonical ensemble, it constitutes an ideal playing ground to test the quality of the approximate transformations eqs.(\ref{eq8}),(\ref{eq5})
in different thermodynamic situations.  If these transformations can be validated in well defined thermodynamic regions
and/or for well defined observables of interest, the natural continuation of this work will be to exploit such transformations to account for situations where no analytical solution exists.
In particular, applying the constraint of energy or angular momentum conservation requires numerically heavy Monte-Carlo techniques with all the associated convergence problems, while an approximate implementation of these conservation laws through appropriate lagrange multipliers (the analogous of the grand canonical ensemble) can be easily implemented.
\begin{figure} [h]
\begin{center}
\includegraphics[width=2.5in,height=3.5in,clip]{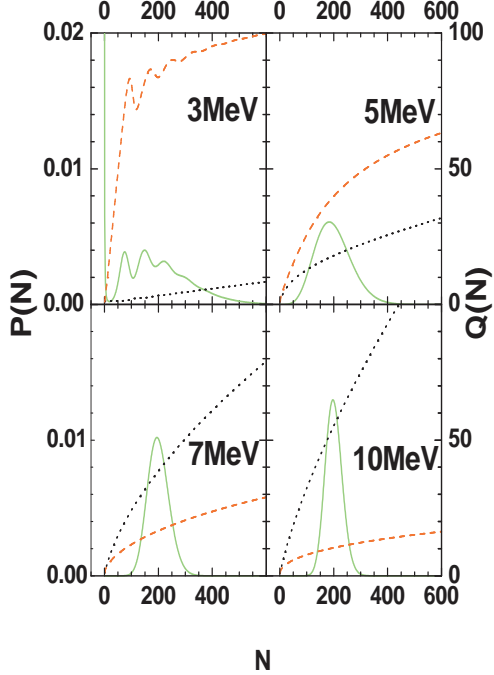}
\caption{  (G)CTM predictions in the finite volume $V=6V_0(200)$  at different temperatures, $T=$ 3, 5, 7 and 10 $MeV$.
Full lines: grandcanonical particle number distributions. Dashed lines: average size of the largest cluster in the canonical ensemble as a function of the total particle number. Dotted lines: average total cluster multiplicity in the canonical ensemble as a function of the total particle number.
}
\end{center}
\end{figure}
Table \ref{table1} shows the performance of eqs.(\ref{eq8}),(\ref{eq5}) for a representative system
with a total number of particles $N=200$  a fixed temperature $T=5$ MeV,
and a fixed volume $V=6V_0(200)$.
These values are  typical for applications to experimental multifragmentation data.

The average multiplicity of a cluster of size $A$ is defined in the two ensembles as\cite{dasgupta}:

\begin{equation}
\langle n\rangle^{A}_{GC}=  \omega_A \exp \alpha_A; \; \;
\langle n\rangle^{A}_{C}=  \omega_A \frac{Z_{N-A}}{Z_N}
\end{equation}

while the total multiplicity is obtained by summing up all the multiplicities of the different sizes.
The average size of the largest cluster can be computed  in the two ensembles as\cite{dasgupta}

\begin{equation}
\langle A_{max}\rangle_{GC}=  \sum_{A=1}^\infty A \left ( 1 - e^{-\langle n\rangle^{A}_{GC}}\right )\prod_{A'>A}
 e^{-\langle n\rangle^{A'}_{GC}}
\end{equation}

and

\begin{equation}
\langle A_{max}\rangle_{C}=  \sum_{A=1}^\infty A \frac{\tilde{Z}_N^{(A)}-\tilde{Z}_N^{(A-1)}}{\tilde{Z}_N}.
\end{equation}

In this last expression,  $\tilde{Z}_N^{(A)}$ is the canonical partition sum of $N$ particles where all $\omega_k$ with $k>A$ have been set to zero.
\begin{table}[h]
\begin{center}
\begin{tabular}{|c|c|c|c|c|}
\hline
& \multicolumn{2}{|c|}{Canonical result} & \multicolumn{2}{c|}{Grand canonical result} \\
\cline{2-5}
Observable & Exact & eq.(\ref{eq8}) & Exact & eq.(\ref{eq5}) \\
 & Cano & & GC & \\
\hline
$<n>_{tot}$ & 18.034 & 18.028 & 17.798 & 17.809\\
\hline
$<n>_{A=1}$ & 1.0778 & 1.0774 & 1.0740 & 1.0745\\
\hline
$<n>_{A=50}$ & 0.0200 & 0.0201 & 0.0223 & 0.0222\\
\hline
$<A_{max}>$ & 39.896 & 39.920 & 38.773 & 38.844\\
\hline
\end{tabular}
\end{center}
\caption{The total average multiplicity, multiplicity of monomers, clusters of $A=50$ particles, and average size of the largest clusters for a system of $<N>=200$ nucleons, a volume $V=6V_0(200)$ and a temperature $T=5$ MeV, as calculated in the different ensembles are compared. The approximation eq.(\ref{eq8}) of the canonical result from the grandcanonical ensemble, and the approximation  eq.(\ref{eq5})  of the grandcanonical result from the canonical ensemble are also given. } \label{table1}
\end{table}
 We can see that the predictions of the two ensembles are very close  for the different observables considered. The residual differences
 can be  very well accounted by  the transformation relations among ensembles.
The good performance of eqs.(\ref{eq8}),(\ref{eq5}) can be understood from the inspection of Fig.1. This figure shows the behavior as
 a function of the particle number
of the canonical multiplicity and size of the largest cluster, as well as the grandcanonical particle number distribution. We can see that at $T=5$ MeV the grandcanonical distribution,
though large and non-gaussian as it is expected in the multifragmentation regime, is still a normal distribution and the canonical observables variation is approximately linear in the
 $N$ interval where the distribution is not negligible. The  performance of the transformation formulas  is worse for $A_{max}$ (0.6 \%) than for $<n_{tot}>$ (0.3\%), but this can be understood
from the fact that the difference between the two ensembles is more important for this highly exclusive observable.
 Conversely, at $T=3$ MeV the grandcanonical distribution is strongly deviating from a gaussian, and presents several peaks.
\begin{table}[b]
\begin{center}
\begin{tabular}{|c|c|c|c|c|c|}
\hline
& & \multicolumn{2}{|c|}{Coulomb on} & \multicolumn{2}{c|}{Coulomb off} \\
\cline{3-6}
N=200 & T & Exact & eq.(\ref{eq8}) & Exact & eq.(\ref{eq8}) \\
 & (MeV) & cano & & cano & \\
\hline
& 3 & 3.344 & 3.242 & 1.0718 & 0.211 \\
\cline{2-6}
$<n>_{tot}$ & 5 & 18.034 & 18.028 & 3.748 & 0.543\\
\cline{2-6}
& 7 & 38.648 & 38.647 & 36.103 & 36.127\\
\cline{2-6}
& 10 & 55.174 & 55.176 & 54.325 & 54.324\\
\hline
& 3 & 83.672 & 86.116 & 199.91 & 181.107\\
\cline{2-6}
$<A_{max}>$ & 5 & 39.896 & 39.920 & 191.97 & 183.138\\
\cline{2-6}
& 7 & 16.625 & 16.624 & 20.031 & 20.043\\
\cline{2-6}
& 10 & 10.352 & 10.352 & 10.737 & 10.737\\
\hline
\end{tabular}
\end{center}
\caption{ Total average multiplicity  and average size of the largest clusters for a system of $<N>=200$ nucleons and a volume $V=6V_0(200)$ at different temperatures.  The approximation eq.(\ref{eq8}) of the canonical result from the grandcanonical ensemble is compared to the exact canonical calculation with and without the Coulomb interaction.  } \label{table}
\end{table}
\begin{figure} [t]
\begin{center}
\includegraphics[width=2.8in,height=2.8in,clip]{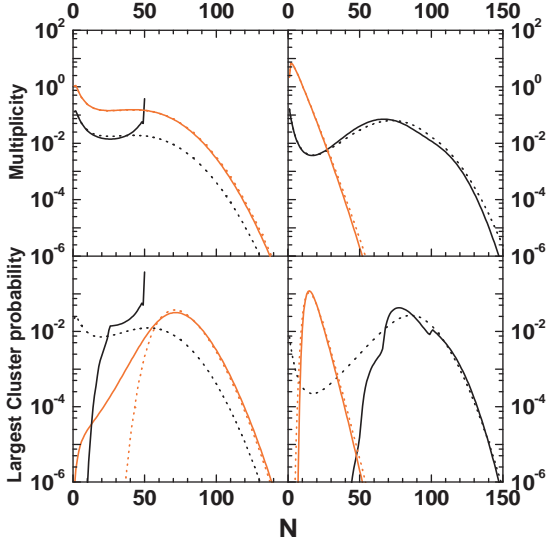}
\caption{  In the upper-left panel and lower left panel  canonical (solid lines) and grand canonical (dotted lines)
 mass distribution and largest cluster probability distribution are shown for A=50 (black) and 400 (red) at T=4 MeV. In the upper-right panel and lower right panel  the same observables are plotted for a system A=200 at T=3 MeV (black) and 7 MeV (red)  ).
 }
 \end{center}
\end{figure}
Indeed at low temperature the equilibrium partitions are dominated by the most bound clusters which lie between $A=75$ and $A=125$ according to the employed liquid drop mass formula. Integer numbers of the most bound clusters therefore maximize the particle number distribution. This effect, combined with the decrease at high $N$ due to the chemical potential constraint, which imposes the average $<N>_{GC}=200$ particle number , and the excluded volume effect, which suppresses the high multiplicity events, leads to the multi-modal shape of the distribution function. As a consequence, we expect eq. (\ref{eq8}) to give a poor approximation of the canonical thermodynamics at $T=3$ MeV. This is confirmed by table 2, which displays the grandcanonical approximation of the canonical ensemble for the chosen observables as a function of the temperature. We can see that the approximation is extremely precise at high temperature, where the distributions are gaussian and the observables linear with the particle number, while larger deviations (3\% for both $<n_{tot}>$ and  $A_{max}$) are observed at $T=3$ MeV.  Similar observations can be drawn from the inspection of table 3, which displays the performance of the grandcanonical estimation eq.(\ref{eq8}) at  fixed temperature $T=4$ MeV as a function of the particle number. The particle number fluctuation increases with decreasing average particle number in the grandcanonical ensemble. As a consequence, the grandcanonical approximation worsens with decreasing $N$, while it is almost perfect for $N=400$.
The canonical and grandcanonical mass and heaviest cluster distributions are directly compared in Fig.2.
\begin{figure} [h]
\begin{center}
\includegraphics[width=2.5in,height=3.5in,clip]{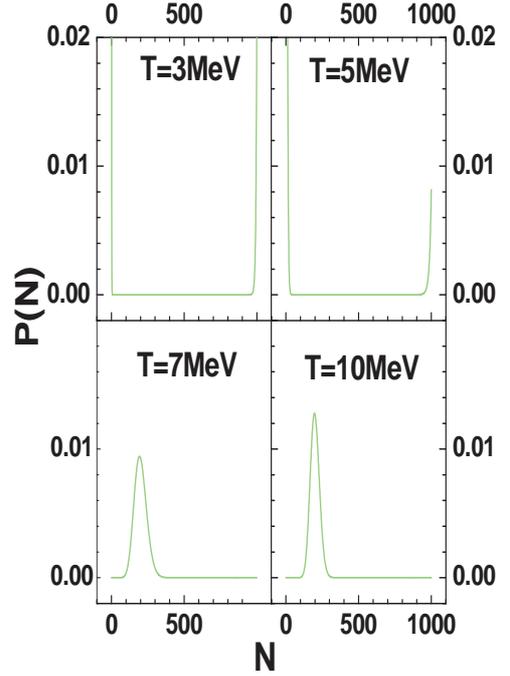}
\caption{  Grand Canonical particle number distributions at different temperatures at $<N_{GC}>=200$ .}
\end{center}
\end{figure}
\begin{table}[b]
\begin{center}
\begin{tabular}{|c|c|c|c|c|}
\hline
T=4 MeV & N & Exact Cano & eq.(\ref{eq8}) & GC  \\
\hline
& 50 & 1.926 & 1.955  & 1.854\\
\cline{2-5}
$<n>_{tot}$ & 100 & 3.859 & 3.808 & 3.708 \\
\cline{2-5}
& 200 & 7.518 & 7.516 & 7.415 \\
\cline{2-5}
& 400 & 14.932 & 14.931 & 14.830 \\
\hline
& 50 & 43.533 & 44.312  & 33.285 \\
\cline{2-5}
$<A_{max}>$ & 100 & 56.997 & 54.877  & 48.442 \\
\cline{2-5}
& 200 & 66.086 & 65.956 & 62.245 \\
\cline{2-5}
& 400 & 73.589 & 73.631  & 72.072 \\
\hline
\end{tabular}
\end{center}
\caption{  Total average multiplicity  and average size of the largest clusters for a system of  volume $V=6V_0(200)$ at  a temperature $T=4$ MeV for different particle numbers.  The grandcanonical result, as well as the approximation eq.(\ref{eq8}) of the canonical result from the grandcanonical ensemble are compared to the exact canonical calculation. }
\end{table}

 We can see that the mass distributions of the two ensembles (upper part) agree reasonably well for all
 temperatures and source mass though the agreement at higher temperatures and masses is definitely much better. Also
 the canonical predictions are trivially
 cut at a size equal to the size of the source.  The close similarity of the distributions explains the
very high accuracy of eq.(\ref{eq8}) concerning multiplicities. The distribution of the heaviest cluster is more interesting.
The canonical distributions are very different from the grandcanonical ones for small systems (lower left)
or low temperatures (lower right). In particular, at temperatures of the order of 3 MeV or lower
 the distribution is clearly bimodal in the grandcanonical ensemble, as expected for a first-order
 liquid-gas phase transition\cite{bimodal_gargi,bimodal_fg}.

It is interesting to remark that bimodal distributions of the heaviest cluster have been reported in experimental fragmentation data\cite{dago,bonnet}. In the case of the experimental samples, the source size is approximately fixed, but since fragmentation occurs in the vacuum the source volume is free to fluctuate. This might allow density fluctuations, similar to our grandcanonical calculations computed in a fixed freeze-out volume. At $T=3$ MeV and for a system of 200 particles, the average size of the largest cluster is $<A_{max}>_{GC}=78.438$, and the performance of  eq.(\ref{eq8}) (see table 2) is remarkable if one considers the huge difference between the distributions shown in Fig.2.

\section{Interpretation and effects of a phase transition}

Globally speaking, these results show
 that the equivalence among the different statistical descriptions is approximately verified, and  our equations are remarkably good in correcting the residual small differences. This is surprising in such small systems, especially considering that the thermodynamic limit of nuclear matter presents a first order phase transition. It is indeed very well known that ensemble inequivalence is especially pronounced in the case of non-extensive systems in the presence of first-order phase transitions\cite{gross,dauxois}.
 In the multi-fragmentation case, no real phase transition can occur because of the Coulomb interaction which prevents  obtaining a thermodynamic limit for the liquid fraction at finite density, and thus quenches the phase transition.
\begin{figure} [h]
\begin{center}
\includegraphics[width=2.8in,height=2.4in,clip]{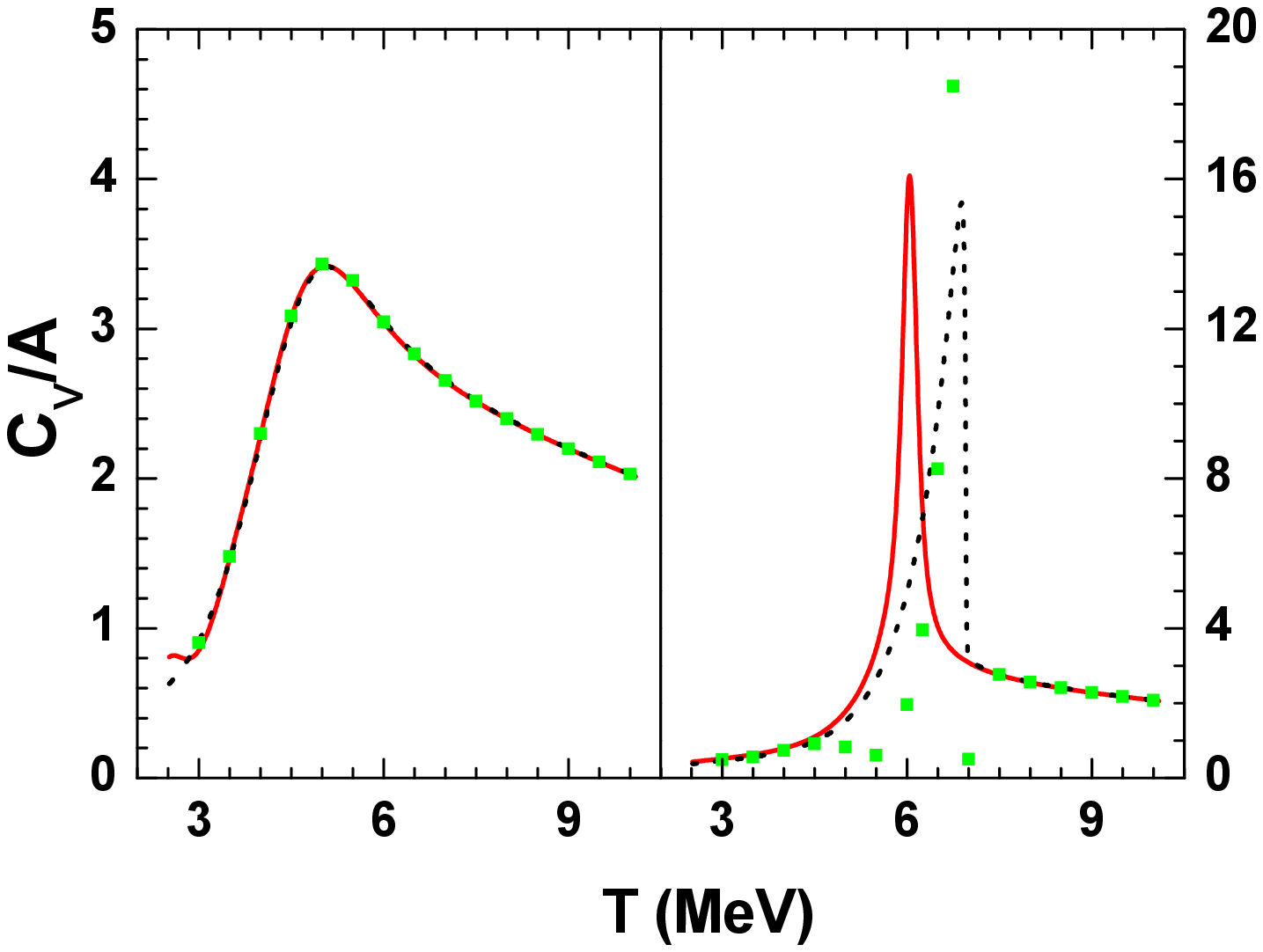}
\caption{  Specific heat per particle as a function of the temperature for a system of 200 particles. Left panel: Coulomb included. Right panel: Coulomb switched off.Full (dashed)  lines: canonical (grand canonical) ensemble. Symbols: estimation of the canonical heat capacity from eq.(\ref{eq8}). }
\label{figcv}
\end{center}
\end{figure}
However the  multiple peaks that we have observed in the particle number distribution at low temperature (Fig.1) and the bimodality of the size distribution of the largest cluster (Fig.2)  are reminiscent of the phase transition of the analogue uncharged system. To show this, we briefly turn to the uncharged case, where the Coulomb energy is artificially switched off. Some selected results are shown in table 2, for the same model cases studied with the full model. We can see
that eq.(\ref{eq8}) badly fails in this case up to a temperature of around $T=5$ MeV. This temperature domain comprises the phase transition, as it can be seen in Fig.3. This figure displays the grandcanonical particle number distribution at different temperatures. The distribution is two-peaked, and the high mass peak corresponds to the maximum cluster size $A_{max}=1000$ which is allowed in the calculation in order to avoid divergencies of the partition sum. This peak physically corresponds to the nuclear liquid fraction, while the peak at $N=1$ corresponds to the nuclear gas fraction.

The effect of a phase transition on the inequivalence between statistical ensembles can be further studied analyzing the isochore heat capacity, which can be straightforwardly calculated from the derivative of the partition sum:
\begin{equation}
c_V=\frac{1}{NT^2} \frac{\partial^2 ln Z}{\partial\beta^2}.
\end{equation}

Results are displayed in Fig.\ref{figcv}.If the Coulomb interaction is included (upper part of Fig.4), the heat capacity presents a large peak that suggests a continuous transition or a cross-over.   Eq.(\ref{eq8}) (symbols in Fig.\ref{figcv}) is very successful in recovering the canonical results from the grand-canonical calculation, but the transformation is useless since the two ensembles produce indistinguishable results.

 If the Coulomb interaction is artificially switched off (lower part of the figure), we can observe that  the discontinuity characterizing a first order phase transition emerges  in the grand canonical ensemble, where finite size effects
are strongly reduced. In this situation ensembles are strongly inequivalent. Finite size effects are very important in the canonical ensemble, leading to a rounding of the transition: the discontinuity is transformed into a peak and shifted towards lower temperatures.
If we try to reconstruct this result employing eq.(\ref{eq8}) (symbols in Fig.\ref{figcv}), we can see that we get a very poor result even at $T=5$ MeV where the two heat capacities are close.
This shows that the failure of the transformation equation is indeed linked to the presence of the phase transition.

\section{Conclusions}

To conclude, in this paper we have analyzed the different sources of non-equivalence between the canonical
 and grandcanonical statistical ensemble in the framework of the (G)CTM model of nuclear multifragmentation.
We have shown that the results of the two ensembles can be transformed into each other with high precision by means of a simple analytical formula. A similar method has
proved to give excellent results in the case of free ideal gases \cite{kosov} and this formula
has been also introduced and used to study statistical ensemble effects in one-dimensional metallic alloys\cite{maras}.
 In this work, we have shown that such analytic expansions do not work when the system experiences a first order phase transition.
 In this case, ensembles are irreducibly non-equivalent and no direct transformation between them is possible.
It is interesting to remark that in the case of phase transitions it is still possible to introduce
intermediate statistical ensembles where fluctuations are constrained, and which continuously interpolate between
canonical and grandcanonical\cite{gaussian}. In the case of nuclear fragmentation though, this is not needed since the
 liquid-gas phase transition is quenched by the Coulomb interaction. As a consequence, the transformation between
the different statistical predictions works remarkably well in the whole thermodynamic region associated
 to the multi-fragmentation phenomenon, even if the ensemble inequivalence associated to the phase transition
 is still visible  in some exclusive observables, notably the distribution of the largest cluster.

\section*{Acknowledgments}
F.G. is indebted to E.Maras and F.Berthier for very stimulating discussions.

\end{document}